\documentclass[sigconf, screen]{acmart}

\usepackage{acronym}
\usepackage{tabularx}
\usepackage{caption}
\usepackage{subcaption}
\acrodef{cps}[CPS]{Cyber-Physical System}
\acrodef{epic}[EPIC]{Electric Power and Intelligent Control}
\acrodef{hai}[HAI]{HIL-based augmented ICS security}
\acrodef{ids}[IDS]{Intrusion Detection System}
\acrodef{ied}[IED]{Intelligent Electronic Device}
\acrodef{ipal}[IPAL]{Industrial Protocol Abstraction Layer}
\acrodef{swat}[SWaT]{Secure Water Treatment}

\colorlet{RED}{red}

\AtBeginDocument{%
  \providecommand\BibTeX{{%
    \normalfont B\kern-0.5em{\scshape i\kern-0.25em b}\kern-0.8em\TeX}}}

\copyrightyear{2022}
\acmYear{2022}
\setcopyright{rightsretained}
\acmConference[CSET 2022]{Cyber Security Experimentation and Test Workshop}{August 8, 2022}{Virtual, CA, USA}
\acmBooktitle{Cyber Security Experimentation and Test Workshop (CSET 2022), August 8, 2022, Virtual, CA, USA}
\acmDOI{10.1145/3546096.3546102}
\acmISBN{978-1-4503-9684-4/22/08}

\begin{document}

%%
%% The "title" command has an optional parameter,
%% allowing the author to define a "short title" to be used in page headers.
\title{PowerDuck: A GOOSE Data Set of Cyberattacks in Substations}

\newcommand{\name}{{\emph{PowerDuck}}}

%%
%% The "author" command and its associated commands are used to define
%% the authors and their affiliations.
%% Of note is the shared affiliation of the first two authors, and the
%% "authornote" and "authornotemark" commands
%% used to denote shared contribution to the research.
%\author{Ben Trovato}
%\authornote{Both authors contributed equally to this research.}
%\email{trovato@corporation.com}
%\orcid{1234-5678-9012}
%\author{G.K.M. Tobin}
%\authornotemark[1]
%\email{webmaster@marysville-ohio.com}
%\affiliation{%
%  \institution{Institute for Clarity in Documentation}
%  \streetaddress{P.O. Box 1212}
%  \city{Dublin}
%  \state{Ohio}
%  \country{USA}
%  \postcode{43017-6221}
%}
  \author{Sven Zemanek}
  \email{sven.zemanek@fkie.fraunhofer.de}
  \affiliation{%
    \institution{Fraunhofer FKIE}
    \country{}
  }
  
  \author{Immanuel Hacker}
  \email{i.hacker@iaew.rwth-aachen.de}
  \affiliation{%
    \institution{RWTH Aachen University}
    \country{}
  }
  \affiliation{%
    \institution{Fraunhofer FIT}
    \country{}
  }
  
  \author{Konrad Wolsing}
  \email{konrad.wolsing@fkie.fraunhofer.de}
  \affiliation{%
    \institution{Fraunhofer FKIE}
    \country{}
  }
  \affiliation{%
    \institution{RWTH Aachen University}
    \country{}
  }

  \author{Eric Wagner}
  \email{eric.wagner@fkie.fraunhofer.de}
  \affiliation{%
    \institution{Fraunhofer FKIE}
    \country{}
  }
  \affiliation{%
    \institution{RWTH Aachen University}
    \country{}
  }
  
  \author{Martin Henze}
  \email{henze@cs.rwth-aachen.de}
  \affiliation{%
    \institution{RWTH Aachen University}
    \country{}
  }
  \affiliation{%
    \institution{Fraunhofer FKIE}
    \country{}
  }

  \author{Martin Serror}
  \email{martin.serror@fkie.fraunhofer.de}
  \affiliation{%
    \institution{Fraunhofer FKIE}
    \country{}
  }

%%
%% By default, the full list of authors will be used in the page
%% headers. Often, this list is too long, and will overlap
%% other information printed in the page headers. This command allows
%% the author to define a more concise list
%% of authors' names for this purpose.
%\renewcommand{\shortauthors}{Trovato and Tobin, et al.}

%%
%% The abstract is a short summary of the work to be presented in the
%% article.
\begin{abstract}
Power grids worldwide are increasingly victims of cyberattacks, where attackers can cause immense damage to critical infrastructure.
The growing digitalization and networking in power grids combined with insufficient protection against cyberattacks further exacerbate this trend.
Hence, security engineers and researchers must counter these new risks by continuously improving security measures.
Data sets of real network traffic during cyberattacks play a decisive role in analyzing and understanding such attacks.
Therefore, this paper presents \name{}, a publicly available security data set containing network traces of GOOSE communication in a physical substation testbed.
The data set includes recordings of various scenarios with and without the presence of attacks.
Furthermore, all network packets originating from the attacker are clearly labeled to facilitate their identification.
We thus envision \name{} improving and complementing existing data sets of substations, which are often generated synthetically, thus enhancing the security of power grids.
\end{abstract}

%%
%% The code below is generated by the tool at http://dl.acm.org/ccs.cfm.
%% Please copy and paste the code instead of the example below.
%%


%%
%% Keywords. The author(s) should pick words that accurately describe
%% the work being presented. Separate the keywords with commas.

%% A "teaser" image appears between the author and affiliation
%% information and the body of the document, and typically spans the
%% page.
%\begin{teaserfigure}
%  \includegraphics[width=\textwidth]{sampleteaser}
%  \caption{Seattle Mariners at Spring Training, 2010.}
%  \Description{Enjoying the baseball game from the third-base
%  seats. Ichiro Suzuki preparing to bat.}
%  \label{fig:teaser}
%\end{teaserfigure}

%%
%% This command processes the author and affiliation and title
%% information and builds the first part of the formatted document.
\maketitle

\section{Introduction}
\label{sec:intro}

The increasing digitalization of power grids holds many opportunities, particularly for shifting from a centralized energy production based on conventional energy sources to a more decentralized production using renewable energies~\cite{KEK+21}.
However, this evolution also amplifies the risks of cyberattacks since initially air-gapped systems become connected to one another and, in some cases, even to the Internet without implementing sufficient security measures~\cite{SHH+21}.
Unfortunately, severe consequences have already manifested in the past in various incidents, such as Stuxnet~\cite{Lang11} or the attacks on the Ukrainian power grid~\cite{WOGS17}.
Hence, academia and industry need to intensify their research in cybersecurity for power grids to effectively improve existing and enable the development of new sophisticated security measures for power grids.

An essential prerequisite for the respective security research is accurate data about the underlying processes, e.g., communication traffic and state information.
Nevertheless, such data is often unavailable to researchers due to security concerns of power grid operators and the high availability requirement of such systems.
Instead, research often relies on artificially generated or simulated data (e.g., \cite{BTZ+19,PFH+19,LFW+21}), subject to abstractions and simplifications.
Meanwhile, a better approach for getting realistic data is performing measurements in a physical testbed with real hardware.

\textbf{Related Work.}
There are already several publicly available data sets recorded in different testbeds and scenarios in the industrial automation domain with an enormous impact on security research.
Examples of such data sets include recordings in the \ac{swat} testbed~\cite{GAJM17} and the \ac{hai} data sets~\cite{SLYK20}.
Possible use cases for these data sets include, for example, the evaluation and enhancement of anomaly detection approaches, such as~\cite{LCJ+19}.
Unfortunately, there is a lack of corresponding real-world security data sets for power grids, and, therefore, researchers often use synthetically generated data to evaluate their approaches (e.g., \cite{MoWa18}).
However, such data is often subject to abstractions and simplifications due to the complex physical processes of power grids and thus has limited use for security research.
Recently, the authors of~\cite{AhKa21} published a data set recorded in the \ac{epic} testbed containing network and process data.
Nevertheless, the considered attack scenarios are limited to malicious reconfigurations of the devices present in the testbed and their impact.
The scenarios in the data set also do not include actual executions of the considered attacks.
Instead, a recommendation is to mutate the collected data, e.g., for anomaly detection, which is again of limited use.

\textbf{Contributions.}
Therefore, this paper presents \name{}\footnote{\url{https://doi.org/10.5281/zenodo.6724225}}, a publicly available data set for cybersecurity research of power grids.
We recorded the data in a real-world testbed representing the secondary technology of a substation in the high and extra-high voltage grid.
The setup consists of three \acp{ied}, a station control system, and the surrounding networking infrastructure.
Moreover, the \acp{ied} also implement a curative measure similar to a busbar protection to represent cascading effects.
In this setting, we recorded the IEC~61850 GOOSE data traffic in different normal operation and attack scenarios with varying parametrizations.
Additionally, we convert our data set to and label malicious packets in the \ac{ipal} format~\cite{wolsing2021ipal} to facilitate their identification and analysis.
The resulting data set thus offers a broad range of possibilities for cybersecurity research and is particularly suited for testing \acp{ids} in power grids.

The structure of this paper is as follows.
Section~\ref{sec:setup} thoroughly describes the real-world testbed we used for the data recordings and the considered operation scenarios.
Then, we provide details on the recorded data in Section~\ref{sec:data}.
We discuss possible use cases of \name{} in Section~\ref{sec:discussion} and conclude this paper in Section~\ref{sec:conclusion}.

\section{Considered Setup}
\label{sec:setup}
In this section, we detail the considered setup for recording \name{}.
We begin with a comprehensive description of the substation testbed in Section~\ref{sec:setup:testbed}.
Then, we present the different scenarios of our measurement campaign in Section~\ref{sec:setup:scenarios}.

\subsection{Testbed}
\label{sec:setup:testbed}
The testbed represents the secondary technology of a substation in the high and extra-high voltage grid, where the setup is sufficiently complex to simulate advanced cyberattacks while still remaining readily understandable.
In the following, we describe the physical structure of the testbed as well as the implemented \ac{ied} application.

\begin{figure}
	\centering
	\begin{subfigure}[b]{0.48\columnwidth}
	   \centering
	   \includegraphics[width=\textwidth]{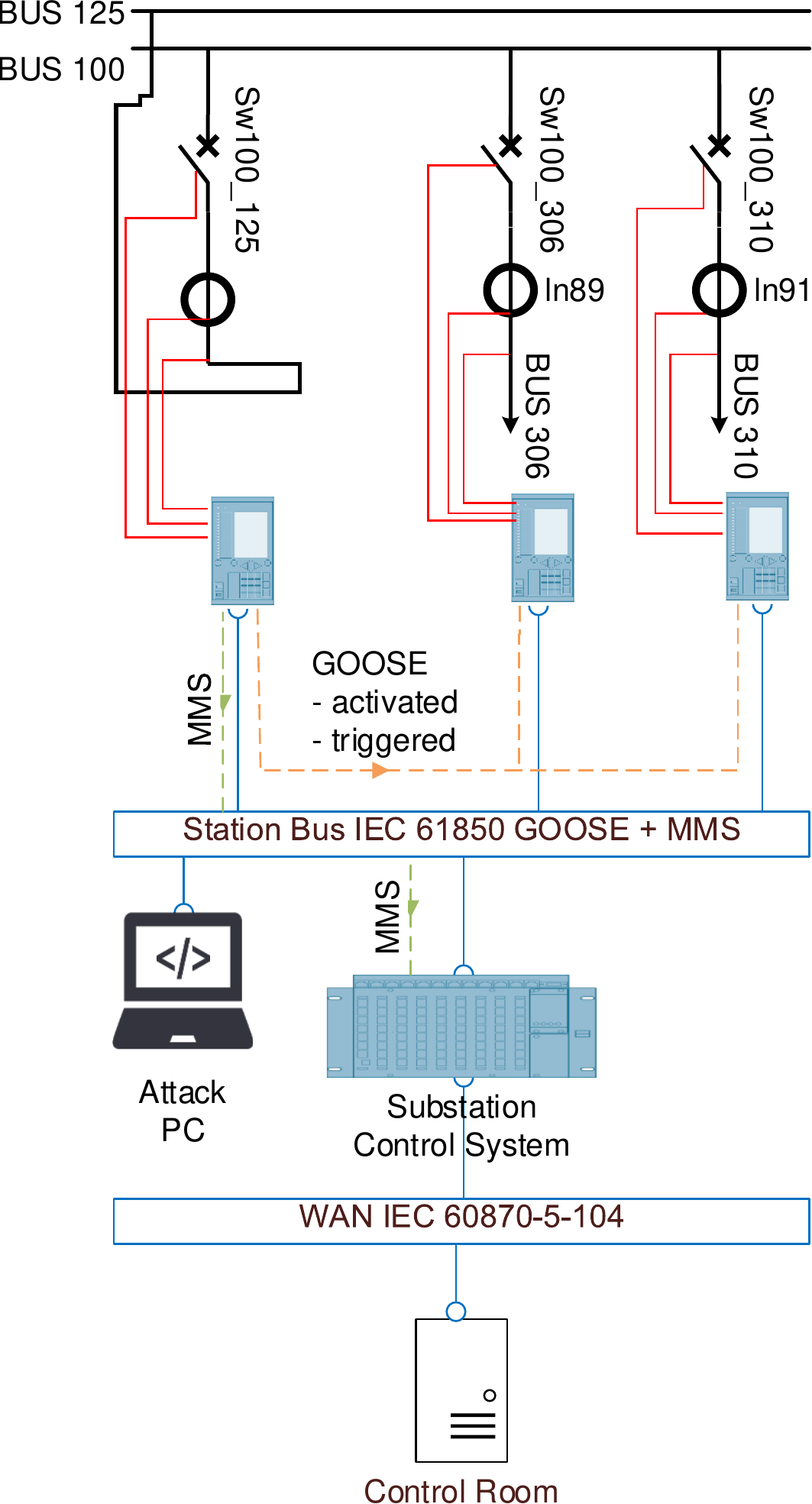}
	   \caption{Schematic view.}
	   \label{fig:setup:schematic}
	\end{subfigure}
	\hfill
	\begin{subfigure}[b]{0.48\columnwidth}
	   \centering
	   \includegraphics[width=\textwidth]{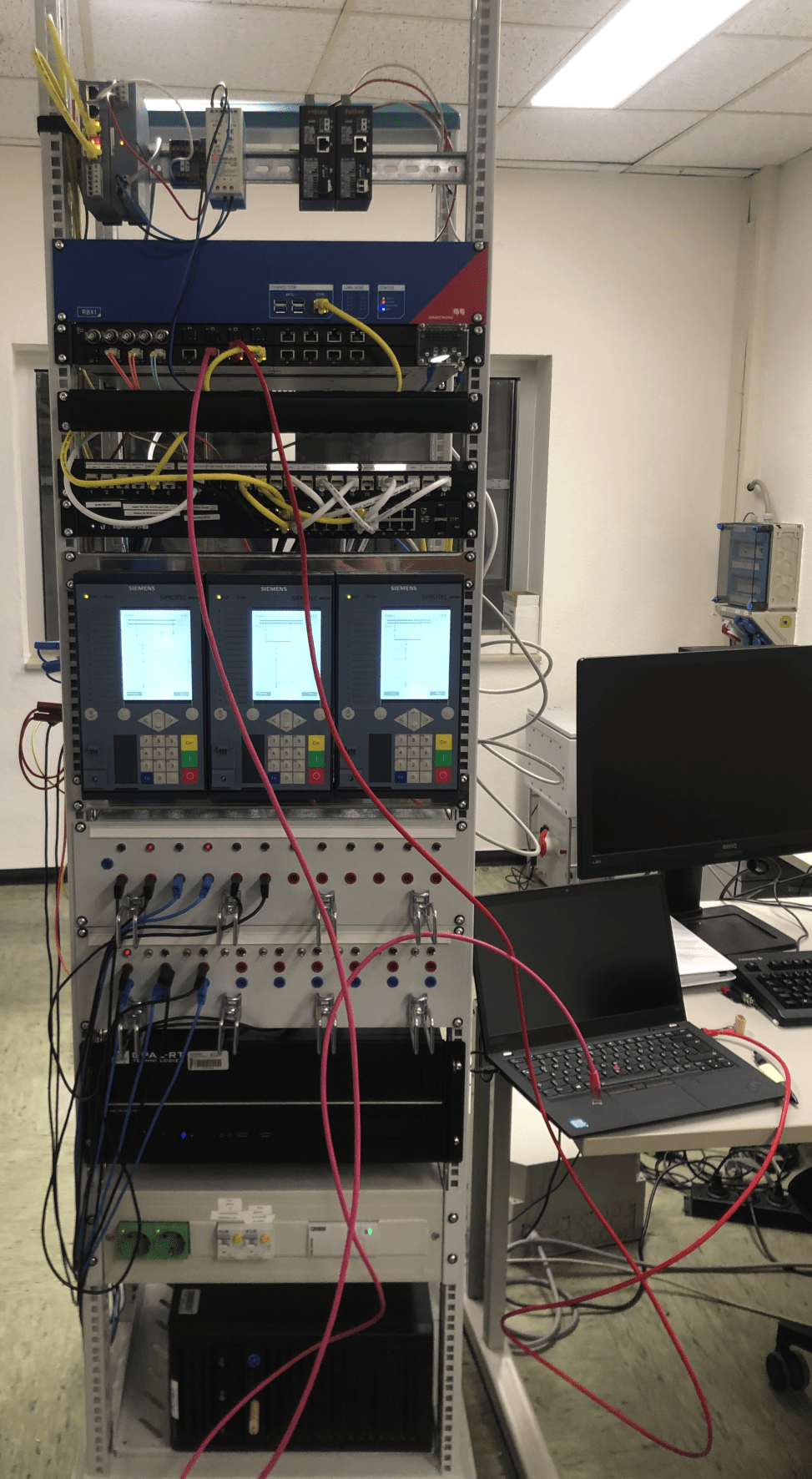}
	   \caption{Photo of the testbed.}
	   \label{fig:setup:photo}
	\end{subfigure}
	\caption{Overview of our substation testbed. The setup consists of a station control system and three IEDs connected through a station bus. Moreover, an attack PC is attached to the station bus, thus with direct access to the local network.}
	\label{fig:setup}
\end{figure}

\paragraph{Structure}
Figure~\ref{fig:setup} depicts our substation testbed.
It consists of a station control system connected to three \acp{ied}.
The communication between the \acp{ied} and the station control system occurs in the same network, which thus replicates the station network of a substation.
To keep the setup manageable, we designed it as a single switch without redundancy systems.
Furthermore, a separate network connects the substation control system with the control room.
The system's scalability is mainly defined by the number of hardware devices and the time to set up the configuration.
In contrast to a fully simulated environment, automatic reconfiguration of the testbed is impossible.
However, it would be possible to extend the testbed by a simulation.

The communication between \acp{ied} and the substation control system uses the IEC~61850 GOOSE protocol and flows through a central switch.
Since an integral part of the \name{} data set consists of traffic recordings, we use the switch's SPAN port to capture all GOOSE packets passing through the switch.
Moreover, as a realistic attack vector in power grids~\cite{KEK+21}, we assume that an attacker controls a device (i.e., a PC) connected to that switch.

For the considered measurement scenarios, we connected a voltage measurement of the public power grid to the three \acp{ied} and implemented a circuit breaker emulation.
This circuit breaker emulation is thus implemented with the help of the I/Os on the \acp{ied} and enables a realistic emulation of the switching process, including correct time delays.
Through this setup, we can verify the effects on the corresponding real-world assets and thus the impact of our attacks.
Mainly, we can observe the opening of circuit breakers when they should not open or vice-versa, and that protection mechanisms, which should open the circuit breakers in case of a fault, do not work.
A result of such attacks could be a partial blackout and even have a more severe impact when executed as a coordinated attack.
Moreover, malfunctioning protection mechanisms may damage equipment and thus lead to prolonged outages.

\paragraph{IED Application}
The \acp{ied} use the IEC~61850 GOOSE protocol to transmit their measurements to the substation control system.
Additionally, the \acp{ied} host an application similar to a busbar protection and implement a curative measure.
This application ensures that when the circuit breaker of the first \ac{ied} is opened, the circuit breakers of the other \acp{ied} are also opened to prevent further damage.
We implemented this behavior using two GOOSE signals, one indicating the activation of the curative measure and the other indicating the current state of the circuit breaker.
Hence, triggering the curative measure would require both signals, an activation and an open circuit breaker.

We defined different measurement scenarios based on this setup to cover distinct testbed behavior under normal and attack conditions.
In the following, we briefly describe these scenarios.

\subsection{Scenarios}
\label{sec:setup:scenarios}

\begin{table*}
\small
\begin{tabularx}{\textwidth}{|p{.01\textwidth}p{.25\textwidth}p{.07\textwidth}p{.07\textwidth}X|}\hline
\textbf{No.} & \textbf{ID} & \textbf{Attack} & \textbf{Duration} & \textbf{Description} \\
\hline
1 & 
normal-1 & 
-- &
00:14:14 &
Open \ac{ied}~1's circuit breaker, then close everything again. \\
\hline
2 & 
normal-2 & 
-- &
00:18:26 &
Open \ac{ied}~$x$'s circuit breaker, then close everything again; for $x$ in \{1,2\}. \\
\hline
3 & 
normal-3 & 
-- &
00:12:38 &
Open \ac{ied}~$x$'s circuit breaker, then close everything again; for $x$ in \{3,2,3\} \\
\hline
4 & 
normal-4 & 
-- &
00:16:53 &
Open \ac{ied}~$x$'s circuit breaker, then close everything again; for $x$ in \{2,1,3,1\}  \\
\hline
\hline
1 &
replay-opening-switch-isolated &
\emph{Replay} &
00:05:35 &
Replay two packets that indicate an opening of \ac{ied}~1's circuit breaker which have been recorded previously. \\
\hline
2 & 
replay-opening-switch-w-context & 
\emph{Replay} &
00:05:36 &
Replay two packets that indicate an opening \ac{ied}~1's circuit breaker and their value repetitions which have been recorded previously. \\
\hline
3 & 
replay-old-measurements & 
\emph{Replay} &
00:05:20 &
Replay previously recorded measurements of \ac{ied}~1.\\
\hline
4 & 
insert-fake-open-w-intermediate & 
\emph{Insertion} &
00:05:07 &
Insert crafted packets (values: opening -- open) that indicate an opening of \ac{ied}~1's circuit breaker. \\
\hline
5 & 
insert-fake-open-only-end & 
\emph{Insertion} &
00:05:24 &
Insert a crafted packet that indicates an open state of \ac{ied}~1's circuit breaker. \\
\hline
6 & 
insert-distort-meas-up-grad &
\emph{Insertion} &
00:10:09 & 
Insert crafted packets that indicate measurements of \ac{ied}~1 raising from 399.45 to 419.91 over 186 seconds. \\
\hline
7 & 
insert-distort-meas-down-grad & 
\emph{Insertion} &
00:10:25 &
Insert crafted packets that indicate measurements of \ac{ied}~1 falling from 399.45 to 380.09 over 176 seconds. \\
\hline
8 & 
insert-distort-meas-up-sharp & 
\emph{Insertion} &
00:11:20 &
Insert a crafted packet that indicates a measurement of \ac{ied}~1 of 420.0. \\
\hline
9 & 
insert-distort-meas-down-sharp & 
\emph{Insertion} &
00:11:32 &
Insert a crafted packet that indicates a measurement of \ac{ied}~1 of 380.0. \\
\hline
10 & 
sup-1-1-tbv0 & 
\emph{Suppression} &
00:10:18 &
Insert two crafted packets (no delay) with a \textit{stNum} value incremented by 1 and 2, respectively, while they still indicate a closed circuit breaker of \ac{ied}~1. \\
\hline
11 & 
sup-1-1-tbv1 & 
\emph{Suppression} &
00:10:24 &
Same as sup-1-1-tbv0, but with 1s delay between the two packets. \\
\hline
12 & 
sup-1-1-tbv2 & 
\emph{Suppression} &
00:12:11 &
Same as sup-1-1-tbv0, but with 2s delay between the two packets. \\
\hline
13 & 
sup-2 & 
\emph{Suppression} &
00:11:29 &
Insert a crafted packet that has a \textit{stNum} value incremented by 2, while its value still indicates a closed circuit breaker of \ac{ied}~1. \\
\hline
14 & 
sup-1 & 
\emph{Suppression} &
00:11:44 &
Insert a crafted packet that has a \textit{stNum} value incremented by 1, while its value still indicates a closed circuit breaker of \ac{ied}~1. \\
\hline
15 & 
flood-repeat & 
\emph{Flooding} &
00:10:34 &
Pick an observed packet announcing the state of \ac{ied}~1 and send it repeatedly in quick succession. \\
\hline
16 & 
flood-bloat-repeat & 
\emph{Flooding} &
00:11:33 &
Same as flood-repeat, but the size of the packet that is being flooded has been increased artificially to about the size of the MTU. \\
\hline
\end{tabularx}
\caption{Description of the measured scenarios in \name{}. For each scenario, the table includes a unique ID to simplify the retrieval of the corresponding data in \name{} and the respective duration (hh:mm:ss) of the measurement.}
\label{tab:scenarios}
\end{table*}

The \name{} data set includes traffic captures from normal operation scenarios as well as different attack scenarios with varying configurations.
The selected attacks represent typical communication attacks targeting the IEC~61850 GOOSE protocol~\cite{BTZ+19} and are thus independent of the used hardware.
As such, they are particularly suited for the \emph{Exploitation} and \emph{Act on Objective} phases of the cyber kill chain model~\cite{YaRa15}.
In the following, we provide a general overview of the considered scenarios for our measurement campaign.
For a detailed description, including the specific parametrizations of the different scenarios, we refer to Table~\ref{tab:scenarios}.

\begin{figure}
\centering
\begin{subfigure}{0.49\columnwidth}
         \includegraphics[scale=0.95]{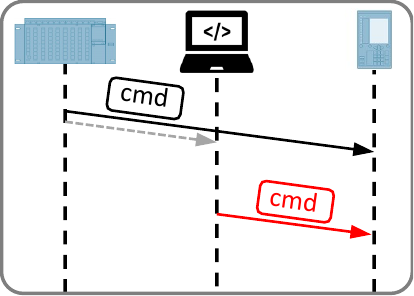}
         \caption{Replay Attack}
         \label{fig:replay}
\end{subfigure}
\hfill
\begin{subfigure}{0.49\columnwidth}
         \includegraphics[scale=0.95]{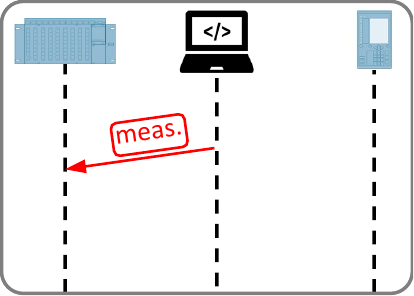}
         \caption{Insertion Attack}
         \label{fig:insertion}
\end{subfigure}
\hfill
\begin{subfigure}{0.49\columnwidth}
         \includegraphics[scale=0.95]{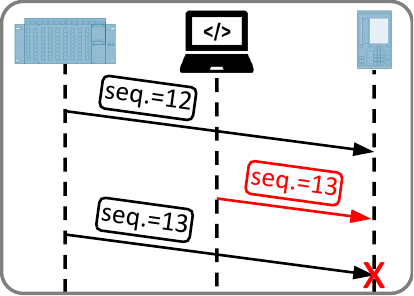}
         \caption{Suppression Attack}
         \label{fig:suppression}
\end{subfigure}
\hfill
\begin{subfigure}{0.49\columnwidth}
         \includegraphics[scale=0.95]{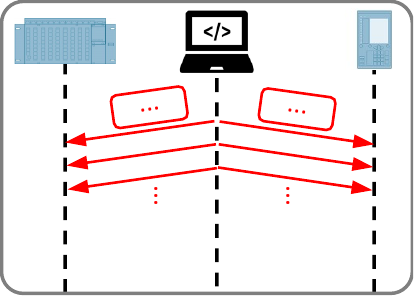}
         \caption{Flooding Attack}
         \label{fig:flooding}
\end{subfigure}
\vspace{-1em}
\caption{Examples for the distinct attacks, including a station control system (left), an attack PC (middle), and an IED.}
\end{figure}

\textbf{Normal Operation.}
The testbed operates normally without any attacks, i.e., the substation control system successfully performs different switching operations.
The recorded traffic from the corresponding four measurement scenarios thus serves as a reference for the normal operation of the involved systems.

\textbf{Replay Attack.}
The attacker eavesdrops on one or more regularly exchanged GOOSE packets to replay them later. 
Such packets could include, e.g., a switching command or an outdated measurement and thus lead to wrong switching operations.
For this attack, we recorded data from three distinct measurement scenarios.
Fig.~\ref{fig:replay} depicts an example of this attack. 

\textbf{Insertion Attack.}
As shown in Fig.~\ref{fig:insertion}, the attacker sends a specially crafted packet to disturb switching operations.
Such a packet could include, e.g., a wrongly announced switch opening or forged measurements, leading to six different scenarios.
For all insertion scenarios, repetition packets were sent in the temporal pattern according to the GOOSE specification.

\textbf{Suppression Attack.}
The attacker exploits GOOSE sequence numbers to manipulate the protocol states of the involved systems, e.g., by sending crafted packets with increasing sequence numbers.
Then, legitimate packets are discarded, which leads to a \emph{denial of service}, as illustrated in Fig.~\ref{fig:suppression}.
We recorded data from five different scenarios.
For all suppression scenarios, repetition packets were sent in the temporal pattern according to the GOOSE specification.

\textbf{Flooding Attack.}
The attacker quickly sends many (large) packets to overburden the communication network, which might lead to packet drops of legitimate requests and hence to a \emph{denial of service}.
We depict the general flooding scenario in Fig.~\ref{fig:flooding} and further recorded two distinct variations of the attack.

With the present description of the testbed setup and the measurement scenarios, we continue with the details of the recording process and the resulting data structure in the following section.

\section{Properties of the Data Set}
\label{sec:data}
This section details how we executed our measurement campaign for \name{} in our real-world testbed using the defined scenarios in Section~\ref{sec:data:collection}. 
Furthermore, we describe the structure of the \name{} data set in Section~\ref{sec:data:structure}.

\subsection{Data Collection}
\label{sec:data:collection}

To capture the relevant network traffic for the different scenarios, we used the SPAN port of the central switch that connects the different entities of our testbed, i.e., the attack PC, the substation control system, and the \acp{ied}.
Moreover, we recorded the outgoing network traffic on the attack PC to unequivocally label data packets originating from the attack PC in the \name{} data set.

We conducted the individual measurements according to their description in Table~\ref{tab:scenarios}.
For each of the $20$~scenarios, we performed the following general steps, where the exact actions depend on the respective parameters and whether it includes an attack or not:

\begin{enumerate}
\item reset the circuit breakers and the substation control system;
\item start the recording of data traffic;
\item possibly start an attack;
\item possibly perform one or more actions, e.g., open a circuit breaker; and
\item stop the recording of data traffic.
\end{enumerate}

The measurement duration varied between $5$ and $18$~minutes, depending on the scenario.
We provide the exact durations in Table~\ref{tab:scenarios}.
Moreover, we slightly varied the timing of the different steps to obtain individual traffic patterns, not depending on a fixed timing.

Upon completing the recording of the data traffic, we performed the following postprocessing steps on the captured network traffic:
We removed all packets from the recordings that do not belong to the IEC~61850 GOOSE protocol.
Furthermore, we identified all packets originating from the attack PC with the help of the attack PC's recordings and described their purpose.
Notably, we distinguished between proper actions and packets that simply repeat previous values according to the IEC~61850 GOOSE specification.

Besides providing the raw packet captures in the PCAP format, we also transcribed them into the \ac{ipal} format~\cite{wolsing2021ipal}, facilitating further analysis and labeling of attack packets.
\ac{ipal} provides an abstract representation of the recorded network packets using the JSON format.
This representation enables a protocol-independent classification of the different packet types and the annotation those packets that originated from the attacker. 

\begin{figure*}
	\centering
	\begin{subfigure}[b]{\textwidth}
		\centering
		\includegraphics[width=\textwidth]{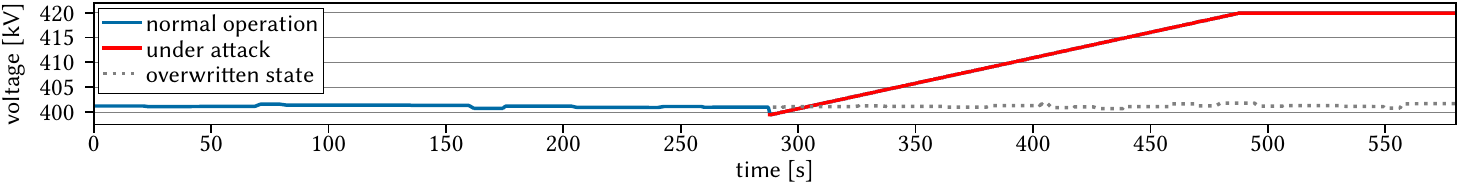}
		\vspace*{-6.5mm}
		\caption{Effects of an insertion attack (no.\ 6) as observed in the \name{} data set.}
		\vspace*{2mm}
		\label{fig:dataset:attack06}
	\end{subfigure}
	\hfill
	\begin{subfigure}[b]{\textwidth}
		\centering
		\includegraphics[width=\textwidth]{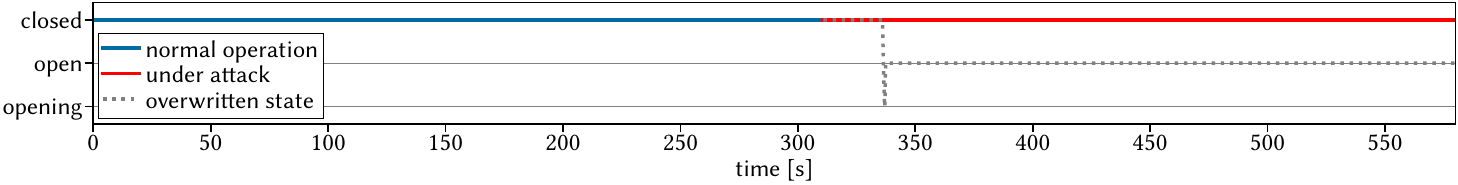}
		\vspace*{-6.5mm}
		\caption{Effects of a suppression attack (no.\ 10) as observed in the \name{} data set.}
		\label{fig:dataset:attack10}
	\end{subfigure}
	\caption{Visualization of two example attacks from the \name{} data set (insertion attack no.~6 and suppression attack no.~10). The plots show the manipulated information along with the actual overwritten states.}
	\label{fig:dataset}
\end{figure*}

As an example, we visualized in Fig.~\ref{fig:dataset} an insertion attack (no.~6) and a suppression attack (no.~10), as observed in the \name{} data set.
The plots show when and how the attacker overwrites the actual process states in the distinct scenarios.
In the following section, we provide further details on the structure of our data set.

\subsection{Data Set Structure}
\label{sec:data:structure}

The \name{} data set contains the recordings of the $20$ previously defined measurement scenarios (Section~\ref{sec:setup:scenarios}).
The data is divided into the plain packet capture files in the \texttt{raw} folder and the \ac{ipal} transcriptions of those packet captures in the \texttt{ipal} folder.
The files in those folders are again subdivided into four files in the \texttt{normal} subfolder, containing regular traffic, and 16 files in the \texttt{attack} subfolder, containing the traffic of the attack scenarios.

In the IPAL transcriptions, packets from regular traffic have their \texttt{malicious} value set to \texttt{false}.
In contrast, packets originating from the attacker have it set to a string indicating the attack ID given in Table~\ref{tab:scenarios}.
This categorization is based on the JSON files in the \texttt{malicious} folder, which indicate and describe the attack packets.
Consequently, such a description does not exist for the normal traffic scenarios, which do not include any attacks.
The files in the \texttt{malicious} folder can be used to identify the attack packets in the packet capture files manually: The \texttt{ipalid} value describes the (0-based\footnote{
	Wireshark displays packet numbers with a 1-based index, e.g., locate the packet with \texttt{ipalid} 200, one has to look for the packet Wireshark displays as No.\ 201.
}) index of the corresponding packet in the packet capture file.
All files in the data set are named according to the scheme \texttt{No-ID-file extension}, where \texttt{No} and \texttt{ID} correspond to those in Table~\ref{tab:scenarios}, e.g., \texttt{01-normal-1.pcapng} for the raw packets captures for the first normal operation scenario.

\section{Applications and Use Cases}
\label{sec:discussion}

After presenting the properties of the \name{} data set in the previous section, this section briefly highlights potential applications and use cases of \name{} for security researchers and engineers.
We thus identify and discuss the following application areas.

\paragraph{Security Analyses}
An essential use of \name{} is to better understand existing attacks on the IEC~61850 GOOSE protocol in the context of power grids and helps to assess potential vulnerabilities.
Such analyses include the concrete manifestation of the attacks on real systems and their possible impact on the network traffic.

\paragraph{(Synthetic) Traffic Generation}
Since some use cases require larger amounts of network traffic than provided by \name{}, a viable approach is to use these real-world recorded traffic patterns to significantly improve synthetic traffic generation, such as proposed in~\cite{SBJ+20}.
Moreover, this approach enables modifications and variations of normal and attack traffic as needed to better fit the requirements of the target use case.

\paragraph{Intrusion Detection Systems}
\name{} offers a solid foundation for the training and testing of \acp{ids} for power grids since the data set also includes labeled packets originating from an attacker.
In particular, the recorded network traffic can be adapted and scaled to match the needed requirements in combination with an improved synthetic traffic generation.\\

To summarize, \name{} offers a solid foundation to address security analyses and implementations in the context of power grids, particularly considering the control in substations via the IEC~61850 GOOSE protocol. 
This real-world data set thus nicely complements and improves existing power grid data sets based on synthetically generated traffic.
Moreover, the considered scenarios include not only the simple transmission of commands but also the (physical) reactions to such controls in an \ac{ied} application, which might trigger further actions and thus influence the network traffic.

\section{Conclusion}
\label{sec:conclusion}

This paper presents \name{}, a publicly available security data set of IEC~61850 GOOSE network traffic in a substation testbed, consisting of a station control system, three \acp{ied}, and an attack PC.
We defined different scenarios, with and without the presence of attacks, to cover a broad behavior of our physical testbed.
Furthermore, the \acp{ied} implement a curative measure, similar to a busbar protection, which enables the triggering of cascading effects.
The resulting data set thus contains the GOOSE network traffic of different scenarios.
Additionally, it includes a transcription to the \ac{ipal} format, which facilitates a protocol-independent analysis and allows us to label the packets originating from the attacker.

Concerning its use cases, \name{} contributes to a deeper understanding of current attacks on substations through the IEC~61850 GOOSE protocol.
More importantly, it offers a solid foundation to complement and improve the synthetic generation of IEC~61850 GOOSE network traffic, as the measurements in \name{} contain actual attacks in a real-world setup.
Hence, it opens up new possibilities for modifying and scaling data sets to individual requirements, such as the training and testing of \acp{ids} in power grids.

%%
%% The acknowledgments section is defined using the "acks" environment
%% (and NOT an unnumbered section). This ensures the proper
%% identification of the section in the article metadata, and the
%% consistent spelling of the heading.
\begin{acks}
This research has been supported by the German Federal Ministry for Economic Affairs and Climate Action within the project ``\mbox{InnoSys 2030} -- Innovations in System Operation up to 2030'' (FKZ: 0350036I).
Only the authors of this publication are responsible for its content.
\end{acks}

%%
%% The next two lines define the bibliography style to be used, and
%% the bibliography file.
\bibliographystyle{ACM-Reference-Format}
\bibliography{paper}

%%% -*-BibTeX-*-
%%% Do NOT edit. File created by BibTeX with style
%%% ACM-Reference-Format-Journals [18-Jan-2012].

\begin{thebibliography}{15}

%%% ====================================================================
%%% NOTE TO THE USER: you can override these defaults by providing
%%% customized versions of any of these macros before the \bibliography
%%% command.  Each of them MUST provide its own final punctuation,
%%% except for \shownote{}, \showDOI{}, and \showURL{}.  The latter two
%%% do not use final punctuation, in order to avoid confusing it with
%%% the Web address.
%%%
%%% To suppress output of a particular field, define its macro to expand
%%% to an empty string, or better, \unskip, like this:
%%%
%%% \newcommand{\showDOI}[1]{\unskip}   % LaTeX syntax
%%%
%%% \def \showDOI #1{\unskip}           % plain TeX syntax
%%%
%%% ====================================================================

\ifx \showCODEN    \undefined \def \showCODEN     #1{\unskip}     \fi
\ifx \showDOI      \undefined \def \showDOI       #1{#1}\fi
\ifx \showISBNx    \undefined \def \showISBNx     #1{\unskip}     \fi
\ifx \showISBNxiii \undefined \def \showISBNxiii  #1{\unskip}     \fi
\ifx \showISSN     \undefined \def \showISSN      #1{\unskip}     \fi
\ifx \showLCCN     \undefined \def \showLCCN      #1{\unskip}     \fi
\ifx \shownote     \undefined \def \shownote      #1{#1}          \fi
\ifx \showarticletitle \undefined \def \showarticletitle #1{#1}   \fi
\ifx \showURL      \undefined \def \showURL       {\relax}        \fi
% The following commands are used for tagged output and should be
% invisible to TeX
\providecommand\bibfield[2]{#2}
\providecommand\bibinfo[2]{#2}
\providecommand\natexlab[1]{#1}
\providecommand\showeprint[2][]{arXiv:#2}

\bibitem[Ahmed and Kandasamy(2021)]%
        {AhKa21}
\bibfield{author}{\bibinfo{person}{Chuadhry~Mujeeb Ahmed} {and}
  \bibinfo{person}{Nandha~Kumar Kandasamy}.} \bibinfo{year}{2021}\natexlab{}.
\newblock \showarticletitle{{A Comprehensive Dataset from a Smart Grid Testbed
  for Machine Learning Based CPS Security Research}}. In
  \bibinfo{booktitle}{\emph{CPS4CIP}}. \bibinfo{publisher}{Springer Int'l
  Pub.}, \bibinfo{address}{Cham}.
\newblock


\bibitem[Biswas et~al\mbox{.}(2019)]%
        {BTZ+19}
\bibfield{author}{\bibinfo{person}{Partha~P. Biswas},
  \bibinfo{person}{Heng~Chuan Tan}, \bibinfo{person}{Qingbo Zhu},
  {et~al\mbox{.}}} \bibinfo{year}{2019}\natexlab{}.
\newblock \showarticletitle{{A Synthesized Dataset for Cybersecurity Study of
  IEC 61850 based Substation}}. In \bibinfo{booktitle}{\emph{IEEE
  SmartGridComm}}.
\newblock


\bibitem[Goh et~al\mbox{.}(2017)]%
        {GAJM17}
\bibfield{author}{\bibinfo{person}{Jonathan Goh}, \bibinfo{person}{Sridhar
  Adepu}, \bibinfo{person}{Khurum~Nazir Junejo}, {and} \bibinfo{person}{Aditya
  Mathur}.} \bibinfo{year}{2017}\natexlab{}.
\newblock \showarticletitle{{A Dataset to Support Research in the Design of
  Secure Water Treatment Systems}}. In \bibinfo{booktitle}{\emph{Critical
  Information Infrastructures Security}}. \bibinfo{publisher}{Springer Int'l
  Pub.}, \bibinfo{address}{Cham}.
\newblock


\bibitem[Krause et~al\mbox{.}(2021)]%
        {KEK+21}
\bibfield{author}{\bibinfo{person}{Tim Krause}, \bibinfo{person}{Raphael
  Ernst}, \bibinfo{person}{Benedikt Klaer}, {et~al\mbox{.}}}
  \bibinfo{year}{2021}\natexlab{}.
\newblock \showarticletitle{{Cybersecurity in Power Grids: Challenges and
  Opportunities}}.
\newblock \bibinfo{journal}{\emph{Sensors}} \bibinfo{volume}{21},
  \bibinfo{number}{18} (\bibinfo{year}{2021}).
\newblock


\bibitem[Langner(2011)]%
        {Lang11}
\bibfield{author}{\bibinfo{person}{Ralph Langner}.}
  \bibinfo{year}{2011}\natexlab{}.
\newblock \showarticletitle{{Stuxnet: Dissecting a Cyberwarfare Weapon}}.
\newblock \bibinfo{journal}{\emph{IEEE Security \& Privacy}}
  \bibinfo{volume}{9}, \bibinfo{number}{3} (\bibinfo{year}{2011}).
\newblock


\bibitem[Li et~al\mbox{.}(2019)]%
        {LCJ+19}
\bibfield{author}{\bibinfo{person}{Dan Li}, \bibinfo{person}{Dacheng Chen},
  \bibinfo{person}{Baihong Jin}, {et~al\mbox{.}}}
  \bibinfo{year}{2019}\natexlab{}.
\newblock \showarticletitle{{MAD-GAN: Multivariate Anomaly Detection for Time
  Series Data with Generative Adversarial Networks}}. In
  \bibinfo{booktitle}{\emph{International Conference on Artificial Neural
  Networks}}. Springer, \bibinfo{address}{Cham}.
\newblock


\bibitem[Lin et~al\mbox{.}(2021)]%
        {LFW+21}
\bibfield{author}{\bibinfo{person}{Chih-Yuan Lin}, \bibinfo{person}{August
  Fundin}, \bibinfo{person}{Erik Westring}, {et~al\mbox{.}}}
  \bibinfo{year}{2021}\natexlab{}.
\newblock \showarticletitle{{RICSel21 Data Collection: Attacks in a Virtual
  Power Network}}. In \bibinfo{booktitle}{\emph{IEEE SmartGridComm}}.
\newblock


\bibitem[Moghaddass and Wang(2018)]%
        {MoWa18}
\bibfield{author}{\bibinfo{person}{Ramin Moghaddass} {and}
  \bibinfo{person}{Jianhui Wang}.} \bibinfo{year}{2018}\natexlab{}.
\newblock \showarticletitle{{A Hierarchical Framework for Smart Grid Anomaly
  Detection Using Large-Scale Smart Meter Data}}.
\newblock \bibinfo{journal}{\emph{IEEE Transactions on Smart Grid}}
  \bibinfo{volume}{9}, \bibinfo{number}{6} (\bibinfo{year}{2018}).
\newblock


\bibitem[Perales~Gómez et~al\mbox{.}(2019)]%
        {PFH+19}
\bibfield{author}{\bibinfo{person}{Ángel~Luis Perales~Gómez},
  \bibinfo{person}{Lorenzo Fernández~Maimó}, \bibinfo{person}{Alberto
  Huertas~Celdrán}, {et~al\mbox{.}}} \bibinfo{year}{2019}\natexlab{}.
\newblock \showarticletitle{{On the Generation of Anomaly Detection Datasets in
  Industrial Control Systems}}.
\newblock \bibinfo{journal}{\emph{IEEE Access}}  \bibinfo{volume}{7}
  (\bibinfo{year}{2019}).
\newblock


\bibitem[Serror et~al\mbox{.}(2021)]%
        {SHH+21}
\bibfield{author}{\bibinfo{person}{Martin Serror}, \bibinfo{person}{Sacha
  Hack}, \bibinfo{person}{Martin Henze}, \bibinfo{person}{Marko Schuba}, {and}
  \bibinfo{person}{Klaus Wehrle}.} \bibinfo{year}{2021}\natexlab{}.
\newblock \showarticletitle{{Challenges and Opportunities in Securing the
  Industrial Internet of Things}}.
\newblock \bibinfo{journal}{\emph{IEEE Transactions on Industrial Informatics}}
  \bibinfo{volume}{17}, \bibinfo{number}{5} (\bibinfo{year}{2021}).
\newblock


\bibitem[Shahid et~al\mbox{.}(2020)]%
        {SBJ+20}
\bibfield{author}{\bibinfo{person}{Mustafizur~R. Shahid},
  \bibinfo{person}{Gregory Blanc}, \bibinfo{person}{Houda Jmila},
  {et~al\mbox{.}}} \bibinfo{year}{2020}\natexlab{}.
\newblock \showarticletitle{{Generative Deep Learning for Internet of Things
  Network Traffic Generation}}. In \bibinfo{booktitle}{\emph{IEEE Pacific Rim
  International Symposium on Dependable Computing}}.
\newblock


\bibitem[Shin et~al\mbox{.}(2020)]%
        {SLYK20}
\bibfield{author}{\bibinfo{person}{Hyeok-Ki Shin}, \bibinfo{person}{Woomyo
  Lee}, \bibinfo{person}{Jeong-Han Yun}, {and} \bibinfo{person}{HyoungChun
  Kim}.} \bibinfo{year}{2020}\natexlab{}.
\newblock \showarticletitle{{HAI 1.0: HIL-based Augmented ICS Security
  Dataset}}. In \bibinfo{booktitle}{\emph{USENIX Workshop on Cyber Security
  Experimentation and Test (CSET '20)}}.
\newblock


\bibitem[Whitehead et~al\mbox{.}(2017)]%
        {WOGS17}
\bibfield{author}{\bibinfo{person}{David~E. Whitehead}, \bibinfo{person}{Kevin
  Owens}, \bibinfo{person}{Dennis Gammel}, {and} \bibinfo{person}{Jess Smith}.}
  \bibinfo{year}{2017}\natexlab{}.
\newblock \showarticletitle{{Ukraine Cyber-Induced Power Outage: Analysis and
  Practical Mitigation Strategies}}. In \bibinfo{booktitle}{\emph{IEEE
  Conference for Protective Relay Engineers}}.
\newblock


\bibitem[Wolsing et~al\mbox{.}(2022)]%
        {wolsing2021ipal}
\bibfield{author}{\bibinfo{person}{Konrad Wolsing}, \bibinfo{person}{Eric
  Wagner}, \bibinfo{person}{Antoine Saillard}, {and} \bibinfo{person}{Martin
  Henze}.} \bibinfo{year}{2022}\natexlab{}.
\newblock \showarticletitle{{IPAL: Breaking up Silos of Protocol-dependent and
  Domain-specific Industrial Intrusion Detection Systems}}. In
  \bibinfo{booktitle}{\emph{International Symposium on Research in Attacks,
  Intrusions and Defenses (RAID '22)}}.
\newblock


\bibitem[Yadav and Rao(2015)]%
        {YaRa15}
\bibfield{author}{\bibinfo{person}{Tarun Yadav} {and}
  \bibinfo{person}{Arvind~Mallari Rao}.} \bibinfo{year}{2015}\natexlab{}.
\newblock \showarticletitle{{Technical Aspects of Cyber Kill Chain}}. In
  \bibinfo{booktitle}{\emph{Security in Computing and Communications}}.
  \bibinfo{publisher}{Springer Int'l Pub.}, \bibinfo{address}{Cham}.
\newblock
\showISBNx{978-3-319-22915-7}


\end{thebibliography}

%%
%% If your work has an appendix, this is the place to put it.
%\appendix

\end{document}